\documentstyle[preprint,eqsecnum,aps]{revtex}
\begin{document}

\tightenlines
\draft

\title{Late-time evolution of the Yang-Mills field in the 
 spherically symmetric  gravitational collapse }

\author{Rong-Gen Cai }
\address{Center for Theoretical Physics, Seoul National
       University, Seoul 151-742, Korea}
\author{ Anzhong Wang}
\address{Departamento de F\'\i sica Te\'orica, Universidade do
         Estado do Rio de Janeiro, \\
         Rua S\~ao Franciso Xavier 524, Maracan\~a, 
      20550-013 Rio de Janeiro-RJ, Brazil}

\maketitle

\begin{abstract}
We investigate the late-time evolution  of the Yang-Mills field in the 
self-gravitating backgrounds: Schwarzschild and Reissner-Nordstr\"om 
spacetimes.  The late-time power-law tails develop in the three asymptotic
regions: the future timelike infinity, the future null infinity and
the black hole horizon. In these two backgrounds, however, the late-time
evolution has quantitative and qualitative differences. In the Schwarzschild 
black hole background, the late-time tails of the Yang-Mills field are the 
same as those of the neutral massless scalar field with multipole moment $l=1$.
The late-time evolution is dominated by the spacetime curvature. When 
the background is the Reissner-Nordstr\"om black hole, the late-time tails
have not only a smaller power-law exponent, but also an oscillatory factor.
The late-time evolution is dominated by the self-interacting term of
the Yang-Mills field. The cause responsible for the differences is
revealed.

\end{abstract}

\section{Introduction}

In the process of gravitational collapse, due to the backscattering off 
the spacetime curvature, the perturbations outside a star or a black hole 
will die off in the form of a inverse power-law tail. 
As a feature of the late-time evolution of gravitational
collapse, the power-law tail has  been studied by many authors.

The late-time evolution of a massless scalar field on a fixed Schwarzschild
background was  investigated first by Price \cite{Price}. He found that an 
initially static $l$ pole dies off as $t^{-(2l+2)}$, while it must fall
off as $t^{-(2l+3)}$ if there is no initial $l$ pole but one develops during
the collapse. Here $t$ is the Schwarzschild coordinate time. The linearized
electromagnetic  and  gravitational perturbations also   satisfy the 
Klein-Gordon equation with a somewhat different effective potential. So the
 electromagnetic  and gravitational  perturbations  have the 
similar late-time behavior as that of the scalar field. 
The late-time power-law tail develops not only at timelike infinity, but also
at null infinity and along the event horizon of black holes \cite{Gund}.
Furthermore, the power-law tail occurs even when no horizon is present in the 
background. This implies that the power-law tail should be present
in perturbations of stars, or after the implosion and subsequent 
explosion of a massless field which does not result in the formation
 of black hole. Indeed this has been confirmed numerically in \cite{Ching}.
The late-time behavior
can also be approached by employing the spectral decomposition 
of corresponding Green's functions \cite{Ching,Lea,And}.

Recently,  Hod and Piran have studied the 
late-time behavior of a massless charged scalar field \cite{Hod1,Hod2},
and a massive scalar field \cite{Hod4} in the gravitational collapse.
Some differences of significance have been observed between the massless 
neutral scalar field and charged scalar field.   In particular, they found 
that the late-time tail  of charged scalar field has an extra 
 oscillatory factor along the black hole horizon. Due to the interaction of 
electromagnetic field, the power-law exponents of the late-time tails 
 are smaller than those of neutral scalar field. 
Therefore, they concluded that a charged black hole becomes bald slower
 than a neutral one.  It is of importance to note that, contrary to the 
neutral scalar field, whose late-time evolution  is dominated by the 
 spacetime curvature, the late-time
evolution of charged scalar field is dominated by the electromagnetic 
interaction, an effect in a  flat spacetime.

According to the no hair theorem of black holes, the collapse of a missive
body may lead to the formation of a black hole and the external gravitational
field of the black hole settles down to the Kerr-Newmann family,
 which characterized by only three parameters: mass, charge and angular 
momentum. Indeed, it has been proved that there do not exist nontrivial
neutral or charged scalar field outside black holes \cite{Bek}. In this sense,
 the late-time power-law tail of scalar fields in fact shows a dynamical 
mechanism by which the scalar fields are  radiated away in the
 gravitational collapse, or perturbations die off. In addition, the form 
of the late-time tail is closely revelant to the internal structure of 
black holes. The late-time tail will act as an input in the study of 
evolution inside black holes. For instance,  the late-time tail
must be used in the mass inflation scenario \cite{Possion}
and in the study of Cauchy horizon stability of black holes.

In the present paper we would like to study the late-time behavior of
the Yang-Mills field in the self-gravitating background.
 Since the discovery of the particle-like solution  by Bartnik 
and McKinnon \cite{BM}, the Einstein-Yang-Mills system (and its 
generalizations) has drawn a great deal of 
interest. In particular, the so-called colored black hole has been found 
\cite{Bizon}, which violates the no-hair theorem of black holes. That is, the
Yang-Mills (YM) field can be regarded as a kind of hairs of black holes.
In addition, the Yang-Mills field has a self-interacting term. We expect 
that it may  give  rise to some interesting phenomena.

The plan of this paper is as follows. For completeness, in the next 
section we briefly introduce the method of spectral decomposition of 
Green's function. It already proves that the Green's function technique is
a  powerful tool to study the dynamical evolution of fields. 
 In Sec. III we linearize the equations 
of motion for the  Einstein-Yang-Mills system and obtain the linearized 
equation of the Yang-Mills field. In Sec. IV and V we study
 the late-time behavior of the Yang-Mills field in  the
Schwarzschild and Reissner-Nordstr\"om black hole backgrounds,
respectively.  Our main results are summarized in  Sec. VI.

\section{Spectral decomposition of evolution fields}

Consider a perturbation field denoted by $\Phi$, which satisfies 
the following equation
\begin{equation}
\label{sec1}
[\partial^2_t -\partial ^2_y +V(y)]\Phi (y,t)=0.
\end{equation}
In order to analytically study the dynamical evolution of the field $\Phi$
in the potential $V(y)$, 
it is convenient to use the Green's function techniques. The evolution of 
$\Phi$ can be determined  by the Green's function and initial conditions as
\begin{equation}
\label{sec2}
\Phi (y, t)=\int [G(y,x;t)\partial _t \Phi (x,0) +\partial _t G(y,x;t)
         \Phi (x,0)]dx,
\end{equation}
for $t\ge 0$. The retarded Green's function $G(y,x;t)$ obeys the
equation
\begin{equation}
\label{sec3}
[\partial^2_t -\partial ^2_y +V(y)]G(y,x;t)=\delta(t)\delta(y-x).
\end{equation}
subject to the  condition  $G(y,x;t)=0 $ for $t\le 0$. In order
to get the Green's function, one may use the Fourier transform
\begin{equation}
\label{sec4}
\tilde{G}(y,x;\sigma)=\int ^{\infty}_{0^-}G(y,x;t)e^{i\sigma t}dt.
\end{equation}
This Fourier transform  is well-defined  in the upper half $\sigma$ plane 
and the Green's function $\tilde{G}(y, x;\sigma)$ satisfies
\begin{equation}
\label{sec5}
[\partial ^2_y +\sigma ^2 -V(y)]\tilde{G}(y,x;\sigma)=\delta(y-x).
\end{equation}
Thus, once given the Green's function $\tilde{G}(y,x;\sigma)$, one can 
obtain the Green's function $G(y,x;t)$ using the inversion transform
\begin{equation}
\label{sec6}
G(y,x;t)=\frac{1}{2\pi}\int^{\infty +ic}_{-\infty+ic}
       \tilde{G}(y,x;\sigma) e^{-i\sigma t} d\sigma,
\end{equation}
where $c$ is some positive constant. To construct 
$\tilde{G}(y,x;\sigma)$, one may use two functions $\phi_i$, 
which are two linearly independent solutions to the homogeneous equation
\begin{equation}
[\partial ^2_y +\sigma^2 -V(y)]\phi_i(y,\sigma)=0, \ \ i=1,2.
\end{equation}
These two functions $\phi_i$ must satisfy appropriate boundary conditions. For
asymptotically flat black hole spacetimes, they should have following 
asymptotic behaviors
\begin{equation}
\label{sec8}
\phi_1 (y,\sigma) \sim \left \{
\begin{array}{ll}
e^{-i\sigma y}, & y\rightarrow -\infty, \\
A_{\rm out}(\sigma)e^{i\sigma y}+A_{\rm in}(\sigma)e^{-i\sigma y}, &
     y\rightarrow +\infty.
\end{array} \right.
\end{equation}
and 
\begin{equation}
\label{sec9}
\phi _2(y, \sigma) \sim \left\{
\begin{array}{ll}
B_{\rm out}(\sigma) e^{i\sigma y} +B_{\rm in}(\sigma)e^{-i\sigma y}, &
         y\rightarrow -\infty, \\
e^{i\sigma y}, & y \rightarrow +\infty.
\end{array} \right.
\end{equation}
That is, $\phi_1$ has only a purely ingoing wave crossing the black hole
horizon $(y\rightarrow -\infty)$. According to the coefficients  in 
(\ref{sec8}), the transmission and reflection amplitudes are
\begin{equation}
\label{sec10}
{\cal T}_1(\sigma)=\frac{1}{A_{\rm in}(\sigma)}, \ \ 
  {\cal R}_1(\sigma)=\frac{A_{\rm out}(\sigma)}{A_{\rm  in}(\sigma)}.
\end{equation} 
$\phi_2$ has only a purely outgoing wave at spatial infinity. The 
transmission and reflection amplitudes are
\begin{equation}
\label{sec11}
{\cal T}_2(\sigma)=\frac{1}{B_{\rm out}(\sigma)}, \ \
{\cal R}_2(\sigma)=\frac{B_{\rm in}(\sigma)}{B_{\rm out}(\sigma)}.
\end{equation}
Therefore, while  (\ref{sec10}) gives the absorption coefficient,
 $T(\sigma)=|{\cal T}_1(\sigma)|^2$, of the black hole,  (\ref{sec11}) gives 
the Hawking radiation coefficient of the black hole, $R(\sigma)
=|{\cal R}_2(\sigma)|^2$.

Using the two functions, the Green's function 
$\tilde{G}(y,x;\sigma)$ can be expressed as
\begin{equation}
\label{sec12}
\tilde{G}(y,x;\sigma)=-\frac{1}{W(\sigma)} 
\left \{
\begin{array}{ll}
\phi_1(y,\sigma)\phi_2(x,\sigma), & y <x, \\
\phi_1(x,\sigma)\phi_2(y, \sigma), & y >x,
\end{array} \right.
\end{equation}
where $W(\sigma)$ is the Wronskian of $\phi_i$, defined as
\begin{equation}
\label{sec13}
W(\sigma)=\phi_1(y, \sigma) \partial_y\phi_2(y, \sigma)
   -\phi_2(y, \sigma)\partial _y\phi_1(y,\sigma)=2i\sigma A_{\rm in}(\sigma).
\end{equation}
The Wronskian is independent of $y$.

To get the Green's function $G(y,x;t)$ in (\ref{sec6}), we must choose an 
appropriate integration contour. Usually one may bend the integration 
contour into the lower half of the complex $\sigma$ plane. In this way, one
can isolate the behavior of the Green's function in the different time 
intervals. The Green's function consists of three parts \cite{Lea,And,Hod2}.

(1) {\it Prompt response}. This part comes from the integral along the
 large semi-circle. so it corresponds to the high-frequency response. 
In the high-frequency limit the Green's function becomes the propagator 
in flat spacetime. This means that  the radiation reaches the observer 
directly from the source. This is therefore a short-time response and 
will die off beyond  some time.

(2) {\it Quasinormal modes}. The Green's function $\tilde{G}
(y,x;\sigma)$ has an infinite number of distinct singularities in the 
lower half plan of the complex $\sigma$. These singularities 
correspond to the black hole quasinormal modes and they occur when the 
Wronskian vanishes there.  This part falls off exponentially because 
of ${\rm Im} \sigma <0$ for each mode.

(3) {\it Late-time tail}. Following the quasinormal modes is just 
 the late-time tail. This part is associated with the existence 
 of a branch cut in the solution $\phi_2$ in this complex picture. 
This cut is usually placed along the negative imaginary $\sigma $ axis. 
The contribution of this part to the 
Green's function  comes from the integral around the branch cut. 
As was shown previously,  this part  generally has an inverse power-law 
form in the asymptotically flat spacetimes \cite{Price}.

In this paper we just study the late-time tail of the YM field in its own 
 gravitational background. So in the next section we first obtain the 
linearized YM equation.

\section{Linearized Einstein-Yang-Mills equations}

Consider the gravitational collapse of the Yang-Mills field, whose dynamics 
is governed by the action
\begin{equation}
\label{action}
S=\frac{1}{16\pi}\int d^4x \sqrt{-g}[R-F_{\mu\nu}F^{\mu\nu}],
\end{equation}
where $R$ denotes the curvature scalar and $F_{\mu\nu}$ is the Yang-Mills field
strength defined as $F=dA +A\wedge A $. Here  $A$ is the Yang-Mills potential.
Due to the conformal invariance, the Einstein equations can be written down
as  
\begin{equation}
\label{eq}
R_{\mu\nu}=2 F_{\mu\lambda}F_{\nu}^{\ \lambda}-\frac{1}{2}g_{\mu\nu} 
      F_{\alpha\beta}F^{\alpha\beta},
\end{equation}
and the equation of Yang-Mills field is $D ^*F=0$.  We now consider the
spherically symmetric gravitational collapse. So the line element can be 
written as
\begin{equation}
\label{metric}
ds^2=-e^{\nu}dt^2 +e^{\lambda}dr^2 +r^2d\theta^2 +r^2\sin ^2\theta d\phi^2.
\end{equation}
Correspondingly, we take the following ansatz for the Yang-Mills potential: 
\begin{equation}
A=w \tau_1 d\theta +(\cot \theta \tau_3 +w \tau_2)\sin\theta d\phi. 
\end{equation}
Here $\tau_i$ ($i=1$, $2$, $3$) are standard generators of su(2) Lie algebra. 
$\nu$, $\lambda$ and $w$ are functions of $r$ and $t$,  and, for simplicity,
we have already set the electric components of the Yang-Mills potential vanish.
In the metric (\ref{metric}), the Einstein equations can be simplified to
\begin{eqnarray}
\label{eq1}
&& \lambda' +\nu' =\frac{4}{r}(w'^2 +\dot{w}^2 e^{-\nu+\lambda }), \\
\label{eq2}
&& \dot{\lambda}=\frac{4w'\dot{w}}{r}, \\
\label{eq3}
&& 1-e^{-\lambda}+ \frac{re^{-\lambda}}{2}(\lambda'-\nu')
 =\frac{(1-w^2)^2}{r^2},
\end{eqnarray}
and the equation of the Yang-Mills potential $w$ is
\begin{equation}
\label{eq4}
\ddot{w}e^{\lambda-\nu}+\frac{\dot{\lambda} -\dot{v}}{2}\dot{w}
         e^{\lambda-\nu}-w''-\frac{\nu'-\lambda'}{2}w' 
          -\frac{(1-w^2)w}{r^2}e^{\lambda}=0,
\end{equation}
where a prime represents derivative with respect to $r$ and an overdot stands 
for derivative with respect to $t$. Here we mention that the critical behavior
of the gravitational collapse of the YM field has been studied in
\cite{Chop}, there two distinct critical solutions have been found 
numerically.

To study the late-time behavior of the Yang-Mills field in the process
of the gravitational collapse, we now linearize the Einstein-Yang-Mills 
equations. Suppose that the final static background is described by functions 
 $\nu_0$, $\lambda_0$ and $w_0$, which depend on $r$ only. The functions 
$\nu$, $\lambda$, and $w$ can be expanded as
\begin{equation}
\nu=\nu_0 +\nu_1, \ \ \lambda=\lambda_0 +\lambda_1, \ \ w=w_0 +w_1.
\end{equation}
Thus we obtain the linearized equations
\begin{eqnarray}
\label{eq10}
&& \lambda_1' +\nu'_1=\frac{8w_0'}{r}w_1', \\ 
&& \dot{\lambda}_1=\frac{4w_0'}{r}\dot{w}_1, \\
\label{eq12}
&& \lambda_1'-\nu_1' +(\frac{2}{r}-\lambda_0'+\nu_0')\lambda_1
   +\frac{8e^{\lambda_0}}{r^3}(w_0-w_0^3)w_1=0, 
\end{eqnarray}
and
\begin{equation}
\label{eq13}
w_1''+\frac{\nu_0'-\lambda_0'}{2}w_1' +\frac{w_0'}{2}(\nu_1'-\lambda_1')
 +\frac{(w_0-w_0^3)}{r^2}e^{\lambda_0}\lambda_1
+\frac{1-3w_0^2}{r^2}e^{\lambda_0}w_1 -e^{\lambda_0-\nu_0}\ddot{w}_1
=0.
\end{equation}
Using (\ref{eq10})-(\ref{eq12}), and defining $w_1=e^{-i\sigma t}
e^{(\lambda_0-\nu_0)/4} \phi(r)$, we have 
\begin{eqnarray}
\label{ymp}
\left [ \frac{d^2}{dr^2} \right. 
    &+& \sigma^2 e^{\lambda_0-\nu_0} -\frac{\nu_0''
        -\lambda''_0}{4} -\frac{(\nu_0'-\lambda_0')^2}{16}
      \nonumber \\
    &+& \left.
      \frac{2w_0'^2}{r}(\frac{2}{r}-\lambda'_0+\nu_0') 
 + \frac{8w_0'}{r^3}(w_0-w_0^3)e^{\lambda_0}
    +\frac{(1-3w_0^2)}{r^2}e^{\lambda_0}\right ]\phi (r)=0.
\end{eqnarray}

\section{Late-time tails in the Schwarzschild background}

In the Einstein-Yang-Mills system, there exist 
 two static, spherically symmetric 
black hole solutions: Schwarzschild and Reissner-Nordstr\"om solutions. 
Besides, there is the so-called colored black hole solution. But, 
the latter is dynamically unstable \cite{Zhou}. So it must  decay 
to the Schwarzschild
solution. Therefore as the final fates of the gravitational collapse of
the YM field, the Schwarzschild and Reissner-Nordstr\"om black holes
are two possibilities.  In this section we discuss the case in which the 
final fate of the collapse is the Schwarzschild black hole.

In this case, we have
\begin{equation}
e^{\nu_0}=e^{-\lambda_0}=1-\frac{2m}{r}, \ \  w_0=\pm 1, 
\end{equation}
where $m$ is the mass of the hole. 
Because the late-time behavior of perturbations is determined by the 
backscattering from the asymptotically far region, the late-time behavior 
is dominated by the low-frequency contribution to the 
Green's function. Thus, as long as the observer is situated far from the 
black hole and the initial data has a considerable support only far from
the black hole, the so-called asymptotic approximation is valid \cite{And}.
That is, a large-$r$ (or equivalently, a low-$\sigma$) approximation 
is sufficient  to study the asymptotic late-time behavior of the 
perturbations. Thus expanding  (\ref{ymp}), up to the terms $O(\sigma^2/r)$
and $O(1/r^2)$, yields 
\begin{equation}
\label{sch1}
\left [\frac{d^2}{dr^2}+\sigma^2 +\frac{4m\sigma^2}{r} -\frac{2}{r^2}
      \right] \phi (r)=0.
\end{equation}
 Introducing $\phi(r)=r^2 e^{i\sigma r}\tilde{\phi}(z)$ with $z=-2i\sigma r$,
one may find the equation satisfied by $\tilde{\phi}$
\begin{equation}
\label{con}
\left [z\frac{d^2}{dz^2} +(4-z)\frac{d}{dz}-(2-2im\sigma)\right] 
   \tilde{\phi}(z)=0.
\end{equation}
This  is a confluent hypergeometric equation. It has two linearly independent 
solutions satisfying the requirement to construct the Green's function 
$\tilde{G}(y,x,\sigma)$ in (\ref{sec12}). The two solutions are 
(for asymptotically far region $r >>m $)
\begin{equation}
\label{solu1}
\phi_1 (r, \sigma)=A r^2 e^{i\sigma r}M(2-2im\sigma, 4, -2i\sigma r),
\end{equation}
and
\begin{equation}
\label{solu2}
\phi_2(r, \sigma)=B r^2 e^{i\sigma r} U(2-2im\sigma, 4, -2i\sigma r).
\end{equation}
Here $A$  and  $B$ are two normalization constants, $M(a,b,z)$ and 
$U(a,b.z)$ are two 
linearly independent solutions to the confluent hypergeometric equation
(\ref{con}).

Following \cite{And,Hod2},  for simplicity,
 here we also assume that the initial 
data has a considerable support only inside the observer. Thus the 
branch cut contribution to the Green's function is
\begin{equation}
\label{green}
G(y, x;t)=\frac{1}{2\pi}\int ^{-i\infty}_{0} \phi_1(x, \sigma)
          \left [\frac{\phi_2(y, \sigma e^{2\pi i})}{W(\sigma e^{2\pi i})}
      -\frac{\phi_2(y, \sigma)}{W(\sigma)}\right] e^{-i\sigma t}d\sigma.
\end{equation}
Because  $M(a,b,z)$ is a single-valued function, 
 one has
\begin{equation}
\label{phi1}
\phi_1(r, \sigma e^{2\pi i})=\phi _1 (r, \sigma).
\end{equation}
$U(a,b,z)$ is many-valued function including a branch cut. Using the 
formula
\begin{equation}
U(a,n+1,ze^{2\pi i})=U(a,n+1,z)+2\pi i \frac{(-1)^{n+1}}{n!\Gamma(a-n)}
      M(a,n+1,z),
\end{equation}
where $n$ is an integer, one may find
\begin{equation}
\label{phi2}
 \phi_2 (r, \sigma e^{2\pi i})=\phi_2(r, \sigma) + 
        \frac{i \pi B} {3A \Gamma(-1-2im\sigma)}\phi_1(r, \sigma).
\end{equation}
Substituting (\ref{phi1}) and (\ref{phi2}) into (\ref{sec13}), we have 
\begin{equation}
W(\sigma e^{2\pi i})=W(\sigma).
\end{equation}
Using the fact that $W(\sigma)$ is independent of $y$, one can use the 
large-$r$ limit of $\phi_i(r,\sigma)$ and reach 
\begin{equation}
W(\sigma)= \frac{3i AB \sigma ^{-3}}{4\Gamma(2-2im\sigma)},
\end{equation}
and 
\begin{equation}
\frac{\phi_2(y, \sigma e^{2\pi i})}{W(\sigma e^{2\pi i})}
-\frac{\phi_2(y, \sigma)}{W(\sigma)}
=\frac{i\pi B}{3A \Gamma(-1-2im\sigma)}
  \frac{\phi_1(y,\sigma)}{W(\sigma)}.
\end{equation}
Substituting them into (\ref{green}), we get
\begin{eqnarray}
\label{green1}
G(y,x;t)&=& \frac{2}{9 A^2}\int ^{-i\infty}_0
 \frac{\Gamma(2-2im\sigma)}{\Gamma (-1-2im\sigma)}
     \sigma ^{3}\phi _1(x,\sigma)\phi_1(y,\sigma)e ^{-i\sigma t}d\sigma 
          \nonumber \\
   &\approx &\frac{4i m}{9A^2}\int ^{-i\infty}_0
      \sigma^{4} \phi_1(y, \sigma )\phi_1 (x,\sigma) e^{-i\sigma t}d\sigma.
\end{eqnarray}

(1). {\it Late-time tail at future timelike infinity}. At future 
timelike infinity $i^+$ (where $x$, $y$ $ << t$), we can use the 
$|\sigma |r<<1 $  limit of the solution 
$ \phi _1(r, \sigma )$. According to Eq. (13.5.5) of \cite{book}, one has
\begin{equation}
\phi_1(r, \sigma )\approx A r^2.
\end{equation}
Putting it  into (\ref{green1}), we obtain
\begin{equation}
\label{greent1}
G(y,x;t)=\frac{32\pi m}{3} (xy)^2 t^{-5}.
\end{equation}

(2). {\it Late-time tail at future null infinity}. At future 
  null infinity ${\cal J}^+$, that is, near the region 
 $y-x <<t<< 2y-x$,  one may use the limit 
$|\sigma |x <<1 $ limit of $\phi_1(x,\sigma)$ and the $|\sigma|y >>1$ 
(${\rm Im} \sigma <0$) limit of $\phi_1(y, \sigma)$. Thus one has
\begin{equation}
\label{phix1}
\phi_1(x,\sigma) \approx A x^2,
\end{equation}
and 
\begin{equation}
\phi_1(y, \sigma) \approx \frac{3! A e^{i\sigma y +2im\sigma \ln y }}
            {\Gamma(2+2im\sigma)} 
      e^{-i\pi (2-2im\sigma)} (-2i \sigma )^{-2 +2im\sigma}
    + \frac{3! A e^{-i\sigma y -2im\sigma \ln y}}{\Gamma(2-2im\sigma)}
      (-2i\sigma )^{-2-2im\sigma},
\end{equation}
by using Eq. (13.5.1) of \cite{book}. Substituting them 
into (\ref{green1}), we have 
\begin{equation}
\label{greenn1}
G(y,x;t)=\frac{4m}{3}x^2 (t-y)^{-3} \approx \frac{4m}{3}x^2 u^{-3}.
\end{equation}

(3). {\it Late-time tail along the black hole horizon}. Near the black 
     hole horizon $ H^+$, (\ref{solu1}) does not satisfy 
the equation of the YM field (\ref{ymp}). Considering (\ref{eq13}) and 
(\ref{ymp}), we have a suitable solution
\begin{equation}
\phi_1(y,\sigma) \approx C e^{-i\sigma[y + 2m \ln (y-2m)]},
\end{equation}
where $C$ may depend on $\sigma$. But to match this solution to the solution 
for $y>>m$, $C$ can be taken to be independent of $\sigma$ \cite{Hod2}.
 Using (\ref{phix1}) acts as $\phi _1(x,\sigma)$,
we get
\begin{eqnarray}
\label{greenb1}
G(y,x;t) &=&  \Gamma_0 \frac{32 m}{3}x^2 [t+y + 
        2m \ln (y-2m)]^{-5} \nonumber \\
        &= & \Gamma_0 \frac{32m}{3}x^2 v^{-5},
\end{eqnarray}
where $\Gamma_0$ is a constant.

 Now some remarks are in order. First, we note that the equation
 (\ref{sch1}) is same as the corresponding one for the scalar field
with multipole moment $l=1$. Hence, these late-time behaviors (\ref{greent1}),
(\ref{greenn1}) and (\ref{greenb1}) of the YM field are same as those 
of massless neutral scalar field with $l=1$. For the latter see
 \cite{And,Hod2}. However, here we should point out that there exist some 
differences between  them. For the scalar field in the Schwarzschild 
background, there is a centrifugal barrier term $l(l+1)/r^2$ in the 
effective potential.  
 We are now considering the spherically symmetric excitation of the YM field,
which corresponds to the s-wave of perturbations.
 The term $2/r^2$, corresponding 
to the $l(l+1)/r^2$ term for the scalar field,   in (\ref{sch1})
comes from the self-interacting term of  the YM field, which can be
 seen clearly from (\ref{ymp}). In this sense, the YM field therefore 
falls off faster  the neutral scalar field. Second, when the background is
the particle-like solution or colored black hole, asymptotically 
one has
\begin{equation}
e^{\nu_0} \approx e^{-\lambda_0} =1 -\frac{2m}{r} + O(1/r^2), \ \ 
w_0= \pm 1 + O(1/r).
\end{equation}
In this case, substituting them into (\ref{ymp}) yields a  same equation as
(\ref{sch1}). Therefore in the background of the particle-like solution 
or the colored black hole, the late-time behavior of the YM field 
is the same as  that in the Schwarzschild background.  This is expected 
because the late-time behavior of perturbations  is determined by the nature 
of far region of backgrounds \cite{Gund}.

\section{Late-time tails in the Reissner-Nordstr\"om background}

In this section we discuss the case when the background  is the 
Reissner-Nordstr\"om black hole. In the Einstein-Yang-Mills system, 
it has been shown that the charged, spherically symmetric  black hole 
solution must be the Reissner-Nordstr\"om solution and the regular monopole 
and dyon do not exist. The no-hair theorem therefore holds for the 
charged black hole \cite{Gal}. Thus, in this case  we have 
\begin{equation}
\label{rn}
e^{\nu_0}=e^{-\lambda_0}=1-\frac{2m}{r}+\frac{g^2}{r^4}, \ \ 
w_0=0,
\end{equation}
where $g^2=1$ is the magnetic charge of the solution. Expanding (\ref{ymp}),
in the far region, reduces to 
\begin{equation}
\label{rn1}
\left [\frac{d^2}{dr^2} +\sigma ^2 +\frac{4m\sigma ^2}{r} +\frac{1}{r^2}
        \right ]\phi (r)=0.
\end{equation}
Introducing 
\begin{equation}
\phi (r)=r^{\frac{1}{2}}e^{i( \frac{\sqrt{3}}{2}\ln \,r +\sigma r)}
       \tilde{\phi}(z), \ \   z=-2i\sigma r,
\end{equation}
from (\ref{rn1}) we obtain
\begin{equation}
\left[ z\frac{d^2}{dz^2} + (1 +i\sqrt{3}-z)\frac{d}{dz} 
     -\left (\frac{1}{2} -\frac{i}{2}(4m\sigma -\sqrt{3})\right )
      \right ]\tilde{\phi}(z)=0.
\end{equation}
Once again, this is a confluent hypergeometric equation. Thus we have two 
equations satisfying the requirement to construct the Green's function
\begin{eqnarray}
\label{phi11}
&& \phi_1(r, \sigma )= A r^{\frac{1}{2}}e^{i (\frac{\sqrt{3}}{2} \ln r
     +\sigma r)}
      M[\frac{1}{2}-\frac{i}{2}(4m \sigma -\sqrt{3}),1+i\sqrt{3},
      -2i\sigma r], \\
&& \phi_2(r, \sigma)= B r^{\frac{1}{2}}e^{i(\frac{\sqrt{3}}{2}\ln r
    +\sigma r)} 
      U[\frac{1}{2}-\frac{i}{2}(4m \sigma -\sqrt{3}),1+i\sqrt{3},
      -2i\sigma r].
\end{eqnarray}
For these two solutions, we have 
\begin{eqnarray}
&& \phi_1(r, \sigma e^{2\pi i})=\phi_1(r, \sigma), \\
&& \phi_2(r, \sigma e^{2\pi i})= e^{2\pi \sqrt{3}}\phi_2(r, \sigma)
  +\frac{A}{B} \frac{(1-e^{2\pi \sqrt{3}})\Gamma(-i\sqrt{3})}
      {\Gamma[\frac{1}{2} -\frac{i}{2}(4m \sigma +\sqrt{3})]}
     \phi_1(r, \sigma).
\end{eqnarray}
The Wronskian satisfies 
\begin{equation}
W(\sigma e^{2\pi i})=e^{2\pi \sqrt{3}}W(\sigma).
\end{equation}
Using the asymptotic behaviors of $M(a,b,z)$ and $U(a,b,z)$ as 
$|z|\rightarrow \infty $, we get
\begin{equation}
W(\sigma)=-AB e^{-\frac{\sqrt{3}}{2}\pi-i\sqrt{3}\ln 2}
           \sigma ^{-i\sqrt{3}}
      \frac{\Gamma(1+i\sqrt{3})}{\Gamma[\frac{1}{2}-\frac{i}{2}(4m\sigma
        -\sqrt{3})]}.
\end{equation}
Further we obtain
\begin{equation}
\frac{\phi_2(y, \sigma e^{2\pi i})}{W(\sigma e^{2\pi i})}
-\frac{\phi_2(y, \sigma)}{W(\sigma)}=
          \frac{B}{A}\frac{\Gamma(-i\sqrt{3})}{\Gamma[\frac{1}{2}
     -\frac{i}{2}(4m\sigma +\sqrt{3})]}
      \frac{(e^{-2\pi \sqrt{3}}-1)}{W(\sigma)} \phi_1(y, \sigma).
\end{equation}
Putting it into (\ref{green}) we reach
\begin{eqnarray}
\label{green2}
G(y,x;t) &=& \frac{1}{2\pi A^2}  
          \frac{(1-e^{-2\pi \sqrt{3}})}{e^{-\frac{\sqrt{3}}{2}\pi 
         -i\sqrt{3}\ln 2}} 
        \frac{\Gamma(-i\sqrt{3})}{\Gamma(1+i\sqrt{3})}
       \nonumber \\
   & \times &
      \int ^{-i\infty}_0
    \frac{\Gamma[\frac{1}{2}-\frac{i}{2}(4m\sigma -\sqrt{3})]}
     {\Gamma[\frac{1}{2}-\frac{i}{2}(4m\sigma +\sqrt{3})]}
    \sigma^{i\sqrt{3}}\phi_1(x,\sigma)\phi_1(y, \sigma)e^{-i\sigma t}
     d\sigma  \nonumber \\
   &\approx & \frac{1}{2\pi A^2}  
          \frac{(1-e^{-2\pi \sqrt{3}})}{e^{-\frac{\sqrt{3}}{2}\pi 
         -i\sqrt{3}\ln 2}} 
        \frac{\Gamma(-i\sqrt{3})}{\Gamma(1+i\sqrt{3})}
        \frac{\Gamma(\frac{1}{2}+i\frac{\sqrt{3}}{2})}
          {\Gamma(\frac{1}{2}-i\frac{\sqrt{3}}{2})} 
     \nonumber \\
     &\times & \int^{-i\infty}_0\sigma^{i\sqrt{3}}\phi_1(x,\sigma)
     \phi_1(y,\sigma) e^{-i\sigma t}d\sigma .
\end{eqnarray}

(1). {\it Late-time tail at future timelike infinity}. At the future 
   timelike infinity $i^+$, as in the Schwarzschild background, we can use the 
     limit $|\sigma| x <<1$ and $|\sigma| y<<1$  for $\phi_(x,\sigma)$ and 
     $\phi_2(y, \sigma)$. That is, we can take
\begin{equation}
\label{phi12}
\phi_1(r, \sigma) \approx A r^{\frac{1}{2}}e^{i\frac{\sqrt{3}}{2}\ln r}.
\end{equation}
Substituting it into (\ref{green2}), we find
\begin{equation}
\label{greent2}
 G(y,x;t)=-i \frac{(1-e^{-2\pi \sqrt{3}})\Gamma(-i\sqrt{3})}
         {2\pi e^{-\pi \sqrt{3} -i\sqrt{3}\ln 2}} 
     \frac{\Gamma(\frac{1}{2}+i\frac{\sqrt{3}}{2})}
         {\Gamma(\frac{1}{2}-i\frac{\sqrt{3}}{2})}
        (xy)^{\frac{1}{2}} e^{i\frac{\sqrt{3}}{2}\ln xy}
        t^{-1 -i\sqrt{3}}.
\end{equation}

(2). {\it Late-time tail at future null infinity}. At the future null 
infinity ${\cal J}^+$, we can use the  $|\sigma| x<<1 $ limit for  
$\phi_1(x,\sigma)$, while  the $|\sigma| y>>1 $ limit for
      $\phi_1(y, \sigma)$.  That is, we take (\ref{phi12}) 
 for $\phi_1(x, \sigma)$, and
\begin{eqnarray}
\phi_1(y,\sigma) &=& A y^{\frac{1}{2}} 
        e^{i(\frac{\sqrt{3}}{2}\ln y + \sigma y)}
      \left \{\frac{ \Gamma (1+i\sqrt{3})e^{-i\pi [\frac{1}{2}-\frac{i}{2}
       (4m\sigma -\sqrt{3})]}}
       {\Gamma[\frac{1}{2}+\frac{i}{2}(4m\sigma +\sqrt{3})]}
      (-2i\sigma y)^{-\frac{1}{2}+\frac{i}{2}(4m\sigma -\sqrt{3})} 
      \right. \nonumber \\
    &+& \left. \frac{\Gamma(1+i\sqrt{3})e^{-2i\sigma y}}
      {\Gamma[\frac{1}{2}-\frac{i}{2}(4m\sigma -\sqrt{3})]}
      (-2i\sigma y)^{-\frac{1}{2}-\frac{i}{2}(4m\sigma +\sqrt{3})}
      \right\},
\end{eqnarray}
using Eq. (13.5.1) of \cite{book}. Substituting them into (\ref{green2}),
in this case we obtain
\begin{equation}
\label{greenn2}
G(y,x;t)=-i\frac{(e^{\pi \sqrt{3}}
        -e^{-\pi\sqrt{3}})}{2\pi \sqrt{2} 
          e^{-i\frac{\sqrt{3}}{2}\ln 2} }
    \frac{\Gamma(-i\sqrt{3})\Gamma(1+i\frac{\sqrt{3}}{2})
     \Gamma(\frac{1}{2}+i\frac{\sqrt{3}}{2})}
    {\Gamma(\frac{1}{2}-i\frac{\sqrt{3}}{2})\Gamma(\frac{1}{2}
      +i\frac{\sqrt{3}}{2})}
        x^{\frac{1}{2}}e^{i\frac{\sqrt{3}}{2}\ln x} 
    u^{-\frac{1}{2}-i\frac{\sqrt{3}}{2}}.
\end{equation}

(3). {\it Late-time tail along the black hole horizon}.
        Near the black hole horizon $H^+$, once again, the solution
     (\ref{phi11}) does not satisfy the equation (\ref{ymp}). The appropriate 
     solution should be
\begin{equation}
\phi_1(y, \sigma) \approx C e^{-i\sigma \left[ y + 
       \frac{1}{2\kappa}\ln (y-r_+)\right ]},
\end{equation}
where $r_+$ is the horizon radius and $\kappa$ is the surface gravity 
on the black hole horizon. Taking (\ref{phi12}) as $\phi_1(x, \sigma)$,
we finally obtain
\begin{equation}
\label{greenb2}
G(y,x;t)= -i\Gamma_0 \frac{(1-e^{-2\pi \sqrt{3}})\Gamma(-i\sqrt{3})}
         {2\pi e^{-\pi \sqrt{3} -i\sqrt{3}\ln 2}} 
     \frac{\Gamma(\frac{1}{2}+i\frac{\sqrt{3}}{2})}
         {\Gamma(\frac{1}{2}-i\frac{\sqrt{3}}{2})}
        x^{\frac{1}{2}} e^{i\frac{\sqrt{3}}{2}\ln x}
        v^{-1 -i\sqrt{3}}.
\end{equation}
where $\Gamma_0$ is a constant.

Comparing the late-time tails (\ref{greent2}), (\ref{greenn2}) and
(\ref{greenb2}) in the Reissner-Nordstr\"om black hole background with
those (\ref{greent1}), (\ref{greenn1}) and (\ref{greenb1}) in the
Schwarzschild black hole background, we can find easily that there are 
a lot of differences between them.  First, we notice that the damping 
exponents are different. The damping exponent in the Schwarzschild background 
is always larger than the corresponding one in the 
Reissner-Nordstr\"om background. In this sense, the YM hair  
 decays in the Schwarzschild background faster than in the
 Reissner-Nordstr\"om background. Another important difference is the 
occurrence of an  oscillatory factor in the Reissner-Nordstr\"om background.
This oscillatory factors are all  present for the three late-time tails. 
Note that for charged scalar field, the oscillatory factor occurs only for 
the late-time tail along the black hole horizon. Second, the late-time 
tails of the YM field are also of qualitative differences. It can be observed 
that the late-time tails (\ref{greent1}), (\ref{greenn1}), and 
(\ref{greenb1}) are all proportional to the mass of the black hole.
 This implies that the 
late-time behavior of the YM field is an effect of spacetime curvature in
the Schwarzschild background. However, the late-time tails in the 
Reissner-Nordstr\"om background have nothing to do with the mass or charge
of the hole. Actually, the late-time behavior of YM field in the
 Reissner-Nordstr\"om background is dominated by the self-interaction 
of the YM field,  an effect in a  flat spacetime.  
The causes responsible for the different results are clear. From (\ref{ymp})
it can be seen that the self-interacting term of YM field provides 
 the excitation with a barrier ($2/r^2$) in the Schwarzschild case,
 but with a well ($-1/r^2$) in the Reissner-Nordstr\"om background. 
In fact, it is the difference that makes  the existence  of the
 particle-like solution and colored black hole and nonexistence of the
 regular monopole and dyon  in the Einstein-Yang-Mills theory with the 
su(2) gauge group.

\section{Conclusions}

We have investigated the late-time evolution of the Yang-Mills field
in its own gravitational backgrounds: Schwarzschild and 
Reissner-Nordstr\"om solutions. The Green's functions describing the late-time
tails are calculated at three asymptotic regions:  the future 
timelike infinity $i^+$,  the future null infinity ${\cal J}^+$ and
 the outer horizon  $H^+$ of black holes.  The late-time evolution is different
in the two backgrounds, quantitatively and qualitatively. When the background 
is the Schwarzschild solution, the late-time tails of the YM field are the same as those of neutral  massless scalar field with multipole moment $l=1$. Note 
that the perturbations  considered in this paper are spherically symmetric
excitations. In this sense, the YM 
hairs die off faster than scalar hairs. Note that all late-time tails 
are proportional to the mass of the black hole. Therefore the 
late-time evolution of YM field in the Schwarzschild background is dominated
by  the  spacetime curvature. This is the same as the neutral massless 
scalar field.  However, there still exists a essential difference.
 For the case of scalar field, the centrifugal barrier term $l(l+1)/r^2$ 
in the effective potential is an effect of angular momentum.  In our case, 
the term $2/r^2$, 
which corresponds to the term $l(l+1)/r^2$ of scalar field, in  
(\ref{sch1}) comes from the self-interacting  term of the YM field.

When the background is the Reissner-Nordstr\"om solution, some interesting 
changes occur. The late-time tails (\ref{greent2}), (\ref{greenn2}) and
(\ref{greenb2}) have not only smaller damping exponents,
compared to those in the Schwarzschild background, but also an oscillatory
factor. This implies that the YM hair falls off in the Schwarzschild 
background faster than in the Reissner-Nordstr\"om background. 
This oscillatory factor is present for all three asymptotic region
with different periods. The period is $2\pi/\sqrt{3}$ for the late-time 
tail at the future timelike infinity and along the black hole horizon, and 
$4\pi/\sqrt{3}$ for the tail at the future null infinity. These late-time 
behaviors are dominated by the self-interacting term of the YM field.
 They are  an effect in a flat spacetime. The cause resulting in the
 different results is that the self-interacting  term of YM field
 gives in the effective potential an attractive term for the
 Reissner-Nordstr\"om background, while a repulsive term for the 
Schwarzschild background.

\section*{acknowledgments}

This work was supported in part by the KOSEF through CTP, Seoul National 
University.

\end{document}